\documentclass[aps,pra,reprint,superscriptaddress,groupedaddress,twocolumn]{revtex4-1}
\usepackage[english]{babel}
\usepackage{ucs}
\usepackage[utf8x]{inputenc}
\usepackage{graphicx}
\usepackage{dcolumn}
\usepackage{bm}
\usepackage{bbm}
\usepackage{amsmath}
\usepackage{amssymb}
\usepackage{mathrsfs}
\usepackage{bbold}
\usepackage{empheq}
\usepackage{amsfonts}
\usepackage{color,colordvi}
\usepackage{physics}
\usepackage[colorlinks=true]{hyperref}
\usepackage[capitalize]{cleveref}

\newcommand{\dark}{{\rm dk}}
\newcommand{\ha}{\hat{a}_1} 
\newcommand{\had}{\hat{a}_1^{\dagger}}
\newcommand{\haa}{\hat{a}_2} 
\newcommand{\haad}{\hat{a}_2^{\dagger}}
\newcommand{\hb}{\hat{b}} 
\newcommand{\hbd}{\hat{b}^{\dagger}}
\newcommand{\hc}{\hat{c}} 
\newcommand{\hcd}{\hat{c}^{\dagger}}

\newcommand{\Uad}{\hat{U}_{\rm ad}} 
\newcommand{\Usad}{\hat{V}} 
\newcommand{\dUsad}{\frac{d}{dt}\hat{V}} 
\newcommand{\Ham}{\hat{H}} 
\newcommand{\Hsad}{\hat{H}_{\rm 1, {\rm dsb}}} 
\newcommand{\uA}{u_A} 
\newcommand{\uB}{u_B} 
\newcommand{\uC}{u_C} 
\newcommand{\uWG}{u_{\rm WG}} 
\newcommand{\duA}{\dot{u}_A} 
\newcommand{\duB}{\dot{u}_B} 
\newcommand{\duC}{\dot{u}_C} 

\newcommand{\vac}{\ket*{\rm vac}}

\begin{document}
\title{Shortcuts to adiabaticity in the presence of a continuum: applications to itinerant quantum state transfer}
\author{Alexandre Baksic}
\affiliation{Department of Physics, McGill University, 3600 rue University, Montréal, Quebec H3A 2T8, Canada}
\author{Ron Belyansky}
\affiliation{Department of Physics, McGill University, 3600 rue University, Montréal, Quebec H3A 2T8, Canada}
\author{Hugo Ribeiro}
\affiliation{Department of Physics, McGill University, 3600 rue University, Montréal, Quebec H3A 2T8, Canada}
\author{Aashish A. Clerk}
\affiliation{Department of Physics, McGill University, 3600 rue University, Montréal, Quebec H3A 2T8, Canada}

\begin{abstract}
We present a method for accelerating adiabatic protocols for systems involving a coupling to a continuum, one that cancels both
non-adiabatic errors as well as errors due to dissipation.  We focus on applications to a generic quantum state transfer problem, where the goal
is to transfer a state between a single level or mode, and a propagating temporal mode in a waveguide or transmission line. 
Our approach enables perfect adiabatic transfer protocols in this setup, despite a finite protocol speed and a finite waveguide coupling.  
Our approach even works in highly constrained settings, where there is only a single time-dependent control field available.   
\end{abstract}
\maketitle



\emph{Introduction--  }Adiabatic quantum evolution provides an efficient and robust way to implement a variety of important quantum operations including state transfer~\cite{Parkins1993,Bergmann1998,Greentree2004,Eckert2007,Tian2012,Wang2012a,Wang2012b}, state preparation~\cite{Vitanov1999,Wunderlich2007,Du2010,Chen2010}, and even quantum logic gates~\cite{Averin1998,Zanardi1999,Aharonov2007}.  While such protocols are robust against timing errors, they are necessarily slow, making them vulnerable to dissipation or fluctuations.  There is thus considerable interest in finding ways to accelerate adiabatic protocols, such that fast evolution is possible without significant non-adiabatic errors \cite{TorronteguiReview,Verdeny2014,Polkovnikov2016a,Polkovnikov2016b,Ribeiro2016}.  These techniques are generally referred to as ``shortcuts to adiabaticity'' (STA), and involve modifying control fields to suppress the net effect of non-adiabatic errors \cite{Demirplak2003,Berry2009,Demirplak2008,Chen2011,Ibanez2012,Baksic2016}.  Recent experiments have successfully implemented versions of these strategies \cite{Bason2012,Du2016,Zhang2013,Zhou2016,An2016}.

A key drawback of standard STA approaches is that they require the exact diagonalization of a time-dependent Hamiltonian, making them unwieldy for systems with many degrees of freedom.  They are thus unsuitable for an important class of quantum state transfer problems, where the goal is to transfer an initial state in a localized system having discrete energy levels to a propagating state a continuum such as a waveguide or transmission line (see, e.g.,~\cite{Fleischhauer2000,Duan2003,Pfaff2016}).

In this paper we present a general method for applying STA to the above class of problems. In many cases, it allows one to derive simple closed-form expressions for accelerated pulse sequences, or at worst, requires a very minimal level of numerical effort.   The method is based on first deriving an effective non-Hermitian Hamiltonian, and then constructing dressed-states and modified control sequences that suppress {\it both} non-adiabatic errors (due to finite protocol speed) and ``dissipative" errors (due to the coupling to the continuum).  We apply our technique to two ubiquitous quantum state transfer problems based on STIRAP (stimulated Raman adiabatic passage) \cite{Bergmann1998}, where state transfer between an internal level and a continuum is facilitated by the adiabatic evolution of a system dark state.  Such protocols have been discussed in systems ranging from atomic cavity QED setups \cite{Fleischhauer2000,Duan2003} to optomechanics \cite{Yin2015}.

   Remarkably, we show that our method works even in the highly constrained protocol introduced by Duan et al. \cite{Duan2003}, where there is only a single time-dependent control field in the Hamiltonian.

\begin{figure}[t]
	\begin{center}
		\includegraphics[width=0.8 \columnwidth]{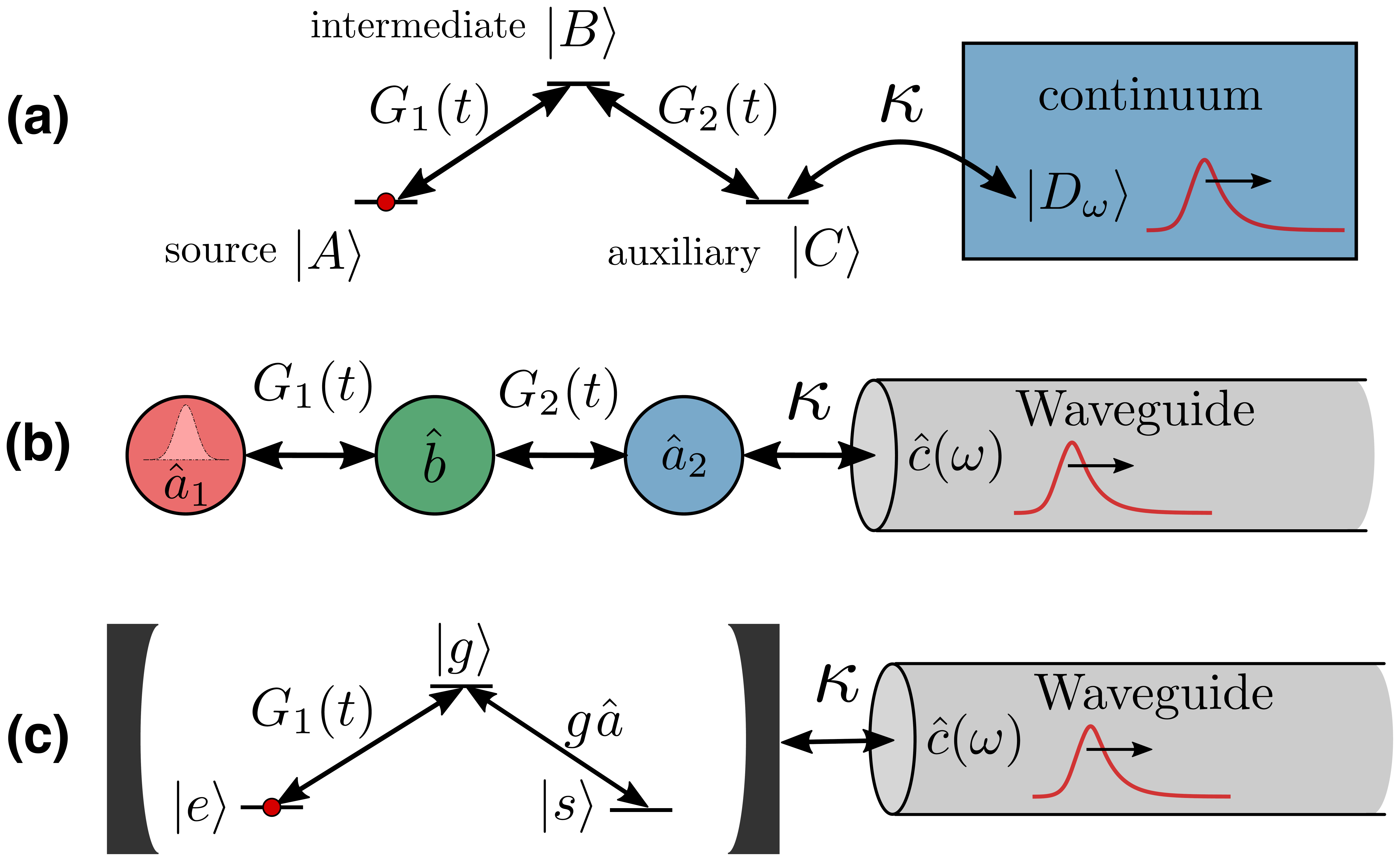}
	\end{center}
	\caption{
	(a)  Three level $\Lambda$ system with time-dependent couplings $G_1(t), G_2(t)$.  The level $\ket*{C}$ is coupled (rate $\kappa$)
	to a waveguide.  The goal is to perform a STIRAP-style state transfer to adiabatically transfer an excitation from $\ket*{A}$ to a propagating
	mode in the waveguide.
	(b)  System of three bosonic modes coupled in a $\Lambda$ configuration, as can be realized in optomechanics \cite{KippenbergRMP2014}
	$\ha$ and $\haa$ are photonic modes, $\hb$ is a mechanical mode, $G_j(t)$ represent many-photon optomechanical couplings.  
	In the single-excitation subspace, this system is completely equivalent to (a).  The correspondence also holds for a general
	initial state due to the linearity of the dynamics, see EPAPS.
	(c)  Realization of (a) using the setup introduced in Refs.~\cite{Fleischhauer2000,Duan2003}, where a three-level system is placed in a cavity (the cavity mode annihilation operator is denoted by $\hat{a}$), which is in turn coupled to a waveguide.  Here, there is only a single time-dependent control field [$G_2(t)=g$ is time-independent].}
\label{fig:sketchSys}
\end{figure}

Our work represents a substantial advance over previous work using STA to accelerate adiabatic state transfer
\cite{Demirplak2003,Berry2009,Demirplak2008,Chen2011,Ibanez2012,Baksic2016}, as these works did not include a coupling to a continuum. It also differs significantly from studies exploring STIRAP-style state transfer to a continuum \cite{Shapiro1994,Vardi1996,Vardi1999,Rangelov2007}, as these did not consider any kind of STA correction.  
Refs.~\cite{Ibanez2011,Torosov2013,Torosov2014} applied STA techniques to phenomenological non-Hermitian Hamiltonians, but in a very different context from the work presented here.

\emph{System--  }
While our approach can be applied to a wide variety of adiabatic protocols [see e.g. Fig.~\ref{fig:sketchSys}(b),(c) and EPAPS], for concreteness we focus on the generic 
state transfer problem depicted in Fig.~\ref{fig:sketchSys}(a), where three discrete levels $A,B$ and $C$ are coupled in a $\Lambda$-system configuration, with the $C$ state 
additionally coupled to a waveguide.  The goal is to convert an initial state $\ket*{A}$ to a propagating excitation in the waveguide.  The system has two time-dependent couplings $G_1(t), G_2(t)$, and is described by the Hamiltonian $\hat{H} = \hat{H}_0(t) + \hat{H}_{\rm int} + \hat{H}_{\rm wg}$, with ($\hbar=1$):  
\begin{align}
	\Ham_{0}(t) =& 
		G_1(t) \dyad*{A}{B}  +G_2(t) \dyad*{C}{B} + \textrm{h.c.}, \nonumber\\
	\Ham_{\rm int} =& \sqrt{\frac{\kappa}{2\pi}}\int_{-\omega_{\rm max}/2}^{\omega_{\rm max}/2}d\omega
	 \Big{[}\dyad*{C}{D_\omega}+\dyad*{D_\omega}{C}\Big{]}, \nonumber\\
	\Ham_{{\rm wg}} = & \int_{-\omega_{\rm max}/2}^{\omega_{\rm max}/2}d\omega\,\, \omega\, \dyad*{D_\omega}{D_\omega}.\label{eq:Hamiltonian_levels_full}
\end{align}
The states in the continuum are defined as $\ket*{D_\omega}=\hcd(\omega)\vac$ where $\hc(\omega)$ is the photon annihilation operator of a mode at frequency $\omega$ in the waveguide, obeying the commutation relation $[\hc(\omega),\hcd(\omega')]=\delta(\omega-\omega')$, and $\vac$ is the vacuum of the whole system. We consider a waveguide with a finite bandwidth $\omega_{\rm max}$, and also that the amplitude of the interaction between the mode $\haa$ and the waveguide is frequency independent [$\kappa(\omega)=\kappa\,\,,\forall\,\,|\omega|\leq\omega_{\rm max}/2$].  

The above model corresponds to the basic setup described in Ref.~\cite{Fleischhauer2000,Duan2003}; we will consider both the case where $G_1(t)$ and $G_2(t)$ are independently tuneable, and the more constrained situation where {\it only} $G_1(t)$ is tuneable.  Note that our results will also immediately apply to the model where the discrete levels $A, B, C$ are replaced by bosonic modes, as is the situation in optomechanical state transfer problems \cite{Wang2012a,Wang2012b,Tian2012}.  In this case, our protocol can be used to transfer an arbitrary $A$-mode state to the state of a propagating wavepacket in the continuum (see EPAPS). 

The starting point for our accelerated adiabatic protocols is the basic STIRAP approach for moving population from $A$ to $C$ in the case where $\kappa=0$ \cite{Bergmann1998}.  This protocol uses the fact that $\hat{H}_0(t)$
has an instantaneous zero-energy eigenstate given by 
\begin{align}
	&\ket*{ {\rm \dark}(t)}=\cos\theta(t)\ket*{A}-\sin\theta(t)\ket*{C}
\end{align}
where we have parameterized the control fields as 
\begin{align}
G_1(t) = G_0(t) \sin \theta(t)\,\,,\,\,G_2(t) = G_0(t) \cos \theta(t).
\end{align}  
This ``dark state" has zero overlap with $\ket*{B}$.  Standard STIRAP \cite{Bergmann1998} works by evolving $\theta(t)$ continuously from $0$ to $\pi/2$, such that  $\ket*{ {\rm \dark}(t)}$ changes continuously from being $\ket*{A}$ at the initial time, to being $\ket*{C}$ at the final time.  If this is done slowly enough compared to the gap $G_0(t)$  separating $\ket*{ {\rm \dark}(t)}$ from the ``bright" adiabatic eigenstates $\ket*{\pm(t)}$,  the system will remain in $\ket*{ {\rm \dark}(t)}$ at all times, thus effecting the desired transfer.

For $\kappa$ non-zero, we could again imagine a STIRAP-like protocol, where the dark state changes adiabatically from being localized in $A$ to $C$.  As $C$ is now coupled to the waveguide, in the ideal case the excitation will be transferred to a propagating waveguide excitation.  This dark state approach for stationary to itinerant state transfer has been discussed in numerous works \cite{Fleischhauer2000,Duan2003,Yin2015}.  STIRAP is an attractive approach as it does not require fine tuning of pulses, and does not involve populating the intermediate state $\ket*{B}$ that might be subject to spurious effects such as e.g. damping and/or dephasing.

\emph{Accelerated STIRAP with dissipation-- }While the above approach has many advantages, any finite speed will lead to non-adiabatic errors which disrupt this transfer.  Ref.~\cite{Baksic2016} presented a dressed-state approach for mitigating this problem in the case where $\kappa=0$.  In our case, the coupling to the waveguide will create additional errors.  We show here how these can also be mitigated using a dressed-state approach.  We start by writing the solution to the Schr\"odinger equation (in the original lab frame) in the form 
\begin{align}
\ket*{\psi(t)} = & \uA(t)\ket*{A} + \uB(t)\ket*{B}+ \uC(t)\ket*{C}\nonumber\\
&+\int_{-\omega_{\rm max}/2}^{\omega_{\rm max}/2}\, d\omega \,\uWG(\omega,t)\ket*{D_\omega} \label{eq:WF_1exc_WG},
\end{align}

One can next solve the linear equations of motion for the waveguide mode amplitudes $\uWG(\omega,t)$, and use these to simplify the equations for the remaining amplitudes.  Taking the Markovian limit where $\omega_{\rm max}\rightarrow\infty$, and assuming that there are no excitations in the waveguide at the initial time $t_i$, one finds that the equations of motion of the remaining amplitudes correspond to a Schr\"{o}dinger equation for the effective non-Hermitian Hamiltonian:
\begin{align}
\Ham_1(t)=\Ham_0(t)-i\frac{\kappa}{2}\dyad*{C}{C}.
\label{eq:Ham_1_phot}
\end{align}

We next transform to the adiabatic frame [via a time-dependent unitary $\hat{U}_{\rm ad}(t)=\sum_{k=\pm,{\rm dk}}\dyad*{k(t)}{k}$], in which the adiabatic eigenstates of $\hat{H}_0(t)$ have no explicit time-dependence.  This involves diagonalizing the three-dimensional Hamiltonian $\Ham_0(t)$ (and not the full, infinite-dimensional Hamiltonian $\Ham$).
In this frame, our effective non-Hermitian Hamiltonian takes the form
\begin{align}
	\Ham_{1,\rm ad}(t)=&  	
				G_0(t)\Big{(}\dyad*{+}{+}-\dyad*{-}{-}\Big{)}-i\frac{\kappa}{2}\sin^2\theta(t)\dyad*{\dark}{\dark}\nonumber\\&-i\frac{\kappa}{2}\cos^2\theta(t)\frac{\ket*{+}+\ket*{-}}{\sqrt{2}}\,\frac{\bra*{+}+\bra*{-}}{\sqrt{2}}\nonumber\\
&-i\left(\dot{\theta}(t)+\frac{\kappa}{4}\sin[2\theta(t)]\right)\frac{\ket*{+}+\ket*{-}}{\sqrt{2}}\bra*{\dark}\nonumber\\&-i\left(-\dot{\theta}(t)+\frac{\kappa}{4}\sin[2\theta(t)]\right)\ket*{\dark}\frac{\bra*{+}+\bra*{-}}{\sqrt{2}}
		\label{eq:H1ad}
\end{align}
The diagonal terms in the first line describe the desired evolution: there is no mixing of the adiabatic eigenstates, and the decay of the dark state corresponds to the desired emission into the waveguide.  The remaining off-diagonal terms describe imperfections.  In particular, both the dissipation ($\kappa \neq 0$) and the finite protocol speed ($\dot{\theta} \neq 0$) cause a deleterious mixing of adiabatic eigenstates.  This implies that, while the deleterious non-adiabatic effects can be arbitrarily reduced by slowing down the protocol, the deleterious effect of the dissipation will still be a problem in this regime.

To design improved pulses that overcome these limitations, we extend the dressed-state approach introduced in Ref.~\cite{Baksic2016}.  
One first constructs a ``dressed" dark state $\ket{\widetilde{\dark}(t)} \equiv \hat{V}(t) \ket{{\dark}}$ that coincides with the original dressed state at the initial and final protocol time.  $\hat{V}(t)$ here is the unitary operator which defines the dressing (in the adiabatic frame).  This dressing reflects the tendency of $\Ham_{1,\rm ad}(t)$ to mix the original adiabatic eigenstates.  Second, one modifies the control pulses $G_1(t), G_2(t)$ such that the dynamics never causes transitions between the dressed dark state and the other two dressed adiabatic eigenstates 
$\ket{\widetilde{\pm}(t)} \equiv \hat{V}(t) \ket{\pm}$.  We describe this modification of the control pulses by an added control Hamiltonian $\Ham_{\rm cor}(t)$, such that the original Hamiltonian is modified as $\Ham_0(t) \rightarrow \Ham_0(t) + \Ham_{\rm cor}(t)$ (in the lab frame).

Formally, the above ``no transitions" requirement is best formulated by writing the effective non-Hermitian Hamiltonian $\Ham_{1,{\rm ad}}(t)$ in the frame where the dressed states $\hat{V}(t) |k \rangle$ have no explicit time dependence (here $k$ labels the original adiabatic eigenstates $+,-,{\dark}$).  This transformed Hamiltonian is given by
\begin{align}
	\Hsad(t) =&\Usad^{\dagger}(t)[\Ham_{{1,\rm ad}}(t)+\Uad^\dagger(t)\Ham_{{\rm cor}}(t)\Uad(t)]\Usad(t)\nonumber\\
	&-i\Usad^{\dagger}(t)\dUsad(t),\label{eq:Hsad}
\end{align}
The requirement that the dynamics does not cause transitions out of the dressed dark state then becomes
\begin{align}
	\bra*{\widetilde{+}}\Hsad(t)\ket*{\widetilde{\dark}}=\bra*{\widetilde{-}}\Hsad(t)\ket*{\widetilde{\dark}}=0.\label{eq:Cancel_off_diag}
\end{align}
Note that as $\Hsad(t)$ is non-Hermitian, fulfilling the above condition does not also imply 
$\bra*{\widetilde{{\dark}}}\Hsad(t)\ket*{\widetilde{\pm}} = 0$.  This is not a concern, as our initial condition (i.e.~we start in the dark state) means that only the matrix elements in Eq.~(\ref{eq:Cancel_off_diag}) are of relevance.  


In order to implement the above strategy, we take a dressing operator $\hat{V}$ whose form corresponds to the kind of mixing of adiabatic eigenstates induced by Eq.~(\ref{eq:H1ad}):
\begin{align}
	&\Usad(t)=\exp\left[i\mu(t) \left(
	\frac{\ket*{+}-\ket*{-}}{\sqrt{2}}\bra*{\dark}+{\rm h.c} \right) \right] .\label{eq:Dress_corr}
\end{align}
Here, $\mu(t)$ parameterizes the strength of the dressing at time $t$.  The fact that the dressing must turn off at the initial and final times
implies that $\mu(t)$ must tend to zero at start and end of the protocol.  

We also parameterize the added correction Hamiltonian via two function $g_x(t)$ and $g_z(t)$:
\begin{align}
\Ham_{\rm cor}(t)=&\Uad(t)\Big{[}g_x(t)\Big{(}\frac{\ket*{+}-\ket*{-}}{\sqrt{2}}\bra*{\dark}+h.c.\Big{)}\nonumber\\
&+g_z(t)\Big{(}\dyad*{+}{+}-\dyad*{-}{-}\Big{)}\Big{]}\Uad^\dagger(t),
\end{align}

which leads modifications of the pulses $G_1(t)$ and $G_2(t)$
\begin{align}
	G_{1,{\rm corr}}(t) &=  G_1(t)-g_x(t)\cos\theta(t)+g_z(t)\sin\theta(t) , \nonumber\\
	G_{2,{\rm corr}}(t) &=   G_2(t)+g_x(t)\sin\theta(t)+g_z(t)\cos\theta(t).
		\label{eqs:GCorrs}
\end{align}
With these definitions in hand, we can now constrain the dressing and modified control pulses so that they fulfil Eq.~(\ref{eq:Cancel_off_diag}), the condition which prevents transitions out of the dressed dark state (either by non-adiabatic errors, or by dissipation).  One finds:
\begin{align}
	&g_x(t)=-\dot{\mu}(t)+\frac{\kappa}{4}\sin^2[\theta(t)]\sin[2\mu(t)]\label{eq:gx}\\
	&g_z(t)=\frac{1}{\tan\mu(t)}\left(\dot{\theta}(t)+\frac{\kappa}{4}\sin[2\theta(t)]\right)-G_0(t),\label{eq:gz}
\end{align}
We thus have an infinite number of corrected protocols that can yield a perfect fidelity despite non-zero $\kappa$ and $\dot{\theta}$:  for any possible dressing function $\mu(t)$ that starts and ends at zero, one simply needs to use modified control pulses that satisfy Eqs.~(\ref{eq:gx})-(\ref{eq:gz}). Useful protocols will correspond to solutions where the modified pulses remain finite in amplitude and close to the original, unmodified pulses.  We discuss two such solutions below.

%
\emph{SATD correction with dissipation-- }
In the case where both $G_1(t)$ and $G_2(t)$ are controllable, one can find a simple correction by choosing the dressing strength $\mu(t)$ so that the control-correction $g_z(t) = 0$.  Using Eq.~(\ref{eq:gz}), we obtain easily:
\begin{align}
	\mu(t)=\arctan\left[  \frac{\dot{\theta}(t)+(\kappa/4)\sin[2\theta(t)]}{G_0(t)}\right].\label{eq:dressing_STIRAP}
\end{align}
Recall that for STIRAP, $\theta(t)$ varies from $0$ to $\pi/2$ during the protocol; hence, the above $\mu(t)$ is guaranteed to vanish at the start and end of the protocol (as required) as long as the original uncorrected protocol is sufficiently smooth.  With $\mu(t)$ determined, the needed modification of the control pulses is given immediately by Eq.~(\ref{eq:gx}) and Eqs.~(\ref{eqs:GCorrs}).

For $\kappa=0$, the dressed states defined by this this choice of $\mu(t)$ corresponds to the instantaneous eigenstates of the adiabatic Hamiltonian $\hat{H}_{\rm 1, ad}$ (the so-called superadiabatic states).  The corresponding corrected pulse sequence then coincides with that described in~\cite{Baksic2016,Zhou2016}.  With non-zero $\kappa$, we see that {\it both} the choice of dressed states and control fields are modified [via the second term in Eq.~(\ref{eq:dressing_STIRAP})].  This modification ensures that irrespective of the size of $\kappa$, we can still have a perfect state transfer from $\ket{A}$ to a propagating temporal mode in the waveguide. We term this new correction scheme ``SATD+$\kappa$". 

With the correction implemented, the dynamics is easy to describe.  One prepares the system in $\ket{A}$ at the initial time $t_i$, which coincides with the dressed dark state, $\ket*{A}$ = $\ket*{\widetilde{\dark}(t_i)}$.  At $t>t_i$, the correction ensures that the system only has amplitude to be in the dressed dark state  $\ket*{\widetilde{\dark}(t)}$ or in the waveguide; the remaining dressed states $\ket{\widetilde{\pm}(t)}$ are never occupied.   Defining $\tilde{u}_{\dark}(t) = \braket*{ \widetilde{\dark}(t)}{ \psi(t)}$, one obtains
\begin{align}
\label{eq:udark-pop}
	|\tilde{u}_{\rm dk}(t)|^2 =& 
		\exp \left[- \int_{t_i}^t  dt' \, \kappa_{\rm eff}(t') \right] \\
	\kappa_{\rm eff}(t') = & \frac{\kappa}{2}\sin^2[\theta(t')]\cos^2[\mu(t')].  
\end{align}  
The physics is thus that the dressed dark state simply leaks directly into the waveguide at an effective instantaneous rate 
$\kappa_{\rm eff}(t)$. The fidelity of the state transfer operation at time $t$, $F(t)$, can then be defined as the probability of having the initial excitation in the waveguide, i.e.
\begin{align}
F(t)=&\int\, d\omega\, |\uWG(\omega,t)|^2=\kappa\int_{t_i}^t\, d\tau\,|\uC(\tau)|^2,
\end{align}
where in the last equality we made use of the expression of $\uWG(\omega,t)$ in the Markovian limit (see EPAPS). A full transfer to the waveguide will thus necessarily require a total protocol time $t_{\rm tot} > 1 / \kappa$.  There is however no additional constraint on the size of the adiabatic gap $G_0(t)$ relative either to protocol time {\it or} the size of the dissipation.

It is also interesting to ask about the temporal mode shape $f(t)$ of the state produced in the waveguide.  This is defined via the amplitude $u_{\rm WG}(\omega,t)$ [c.f.~Eq.~(\ref{eq:WF_1exc_WG})] at the end of the protocol, and is completely determined by the time-dependent amplitude associated with $\ket*{C}$, $u_C(t)$:
\begin{align}
	f(t)=&\lim_{T\rightarrow\infty}\left[\int_{-\infty}^{+\infty}\frac{d\omega}{\sqrt{2\pi}}\exp\left[-i\omega\left(t-T\right)\right]\uWG(\omega,T)\right]\nonumber\\
		=&
		-i\sqrt{\kappa}\,\uC(t),\label{eq:tdef_emp_mode}
\end{align} 

Finally, the perfect fidelity possible with the corrected protocol (irrespective of $\kappa$ and $G_0$) does come with a price: the use of a dressed dark state to effect the transfer means that at intermediate times, the level $\ket{B}$ will have a non-zero occupancy.  The population in this state during the protocol is given by:
\begin{align}
	|\uB(t)|^2=\sin^2[\mu(t)]\exp\left[- \int_{t_i}^t  dt' \, \kappa_{\rm eff}(t') \right].  \label{eq:popB}
\end{align}
As the dressing strength $\mu(t)$ is proportional to $\dot{\theta}(t)$ [see Eq.\eqref{eq:dressing_STIRAP}] , the faster the protocol speed, the greater the population of $\ket*{B}$ at intermediate times.

\begin{figure}[t]
	\begin{center}
		\includegraphics[width=1\columnwidth]{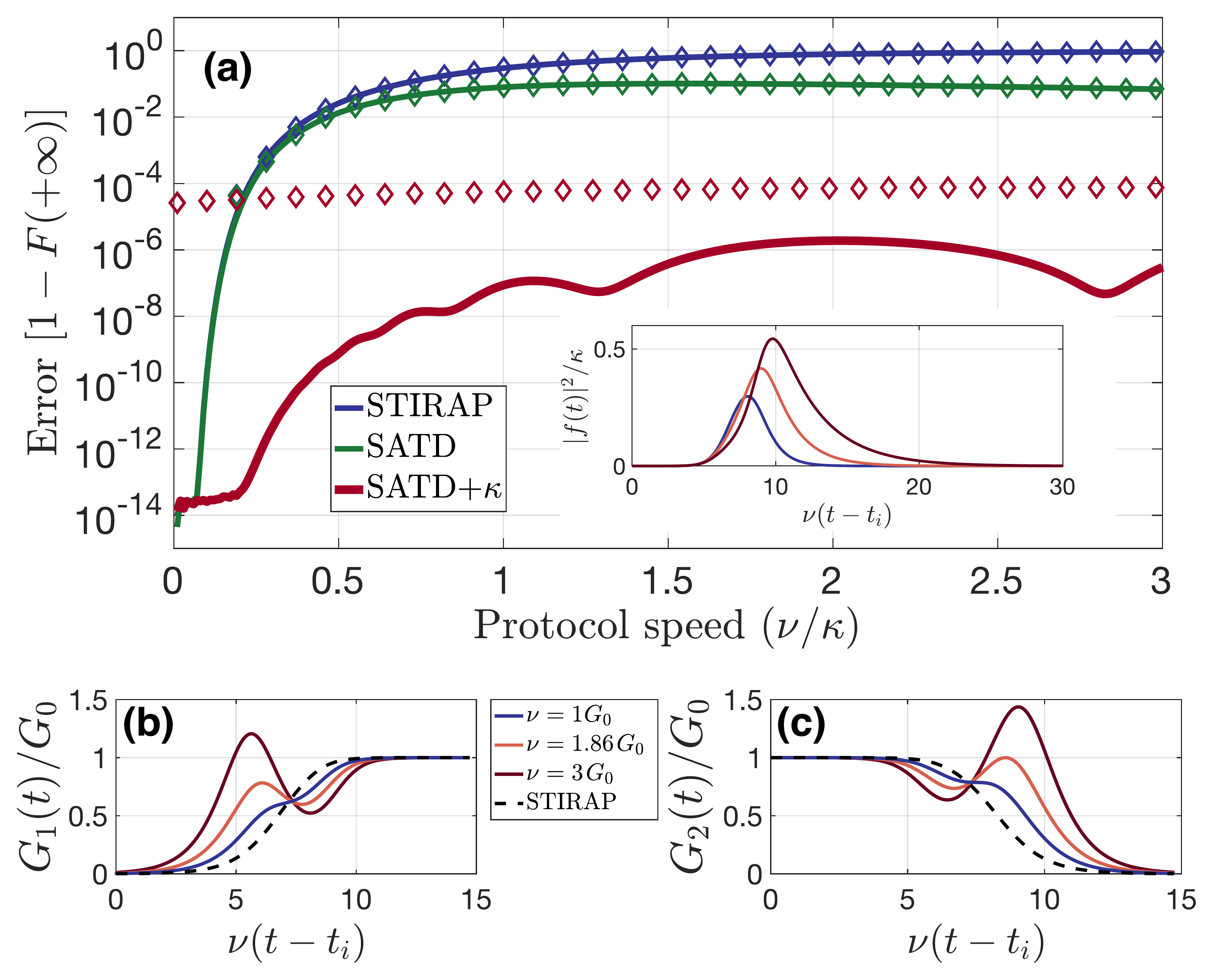}
	\end{center}
	\caption{SATD+$\kappa$ correction for STIRAP-style state transfer to a waveguide based on the optimal 
	STIRAP pulses given in Eq.~(\ref{eq:VitanovPulses}).  
	We take $\kappa$ to be equal to the adiabatic gap $G_0$, a regime in which dissipative errors
	are large.  
	(a)  Asymptotic fidelity $\lim_{t\rightarrow\infty}F(t)$ as a function of protocol speed $\nu$  for the uncorrected
	STIRAP protocol (blue), the SATD protocol (green) and the new SATD+$\kappa$ protocol (red).
	The incoherent decay rate of middle $\ket*{B}$ level is either $\Gamma=0$ (solid curves) or $\Gamma = 10^{-3} \kappa$
	(diamonds).  For $\Gamma = 0$, the fidelity error of the SATD+$\kappa$ protocol is only limited by our truncation of the pulses:  
	initial and final time have been chosen such that $G_1(t_i)=G_2(t_f)= 10^{-3} G_0$. Inset: Shape of the emitted temporal mode in the waveguide when using the corrected pulse sequence [same
	labelling as in (b)].
	(b),(c)  Time-dependence of uncorrected and corrected pulse amplitudes 
	$G_1(t)/G_0$ and $G_2(t)/G_0$ during the protocol.}
	\label{fig:One_exc_comparison}
\end{figure}

To demonstrate the utility of SATD+$\kappa$, we use it to correct the optimal STIRAP pulses
discussed by Vitanov {\it et al.} in Ref.~\cite{Vasilev2009}.  They are defined by
\begin{equation}
 	G_0(t)=G_0,  \hspace{0.2 cm} \theta(t)= \pi / [2 (1+e^{-\nu t} )], 
	\label{eq:VitanovPulses}
 \end{equation}
 and only turn on and off asymptotically as $t \rightarrow \pm \infty$.  To mimic a realistic experiment, 
we truncate the pulses to a finite time interval $-t_i = t_f \simeq 7.4/ \nu$, which ensures $G_1(t_i)=G_2(t_f)= 10^{-3} G_0$.
Fig.~\ref{fig:One_exc_comparison}(a) shows the asymptotic behavior of fidelity $\lim_{t\rightarrow\infty}F(t)$ for this protocol versus the protocol speed $\nu$, with comparisons against both our SATD+$\kappa$ correction, and the $\kappa=0$ correction. The SATD+$\kappa$ correction yields several orders of magnitude improvement.  Note that the only reason it fails to be perfect is due to constraining the pulses to a finite time interval. Moreover, even when we include incoherent decay on the intermediate level at a rate $\Gamma=10^{-3}G_0$ (diamonds), the SATD+$\kappa$ correction still yields several orders of magnitude improvement. Figs.~\ref{fig:One_exc_comparison}(b) and (c) show the form of the corrected pulse sequences, while the inset of (a) shows the final outgoing temporal mode when using the SATD+$\kappa$ correction. 

\emph{Accelerated STIRAP using a single control field-- } Adiabatic state transfer to a waveguide is also possible in systems
where  $G_2(t) = g$ is a fixed constant, 
and only $G_1(t)$ is controllable [e.g.~Fig.~\ref{fig:sketchSys}(b)]  \cite{Fleischhauer2000,Duan2003}.
The SATD+$\kappa$ approach for correcting errors is no longer viable, as it requires both $G_1(t)$ and $G_2(t)$ to be time-dependent.  Nonetheless, by using an alternate form of dressing, we can still obtain a perfectly corrected protocol in this more constrained setting.

\begin{figure}[t]
	\begin{center}
		\includegraphics[width=1\columnwidth]{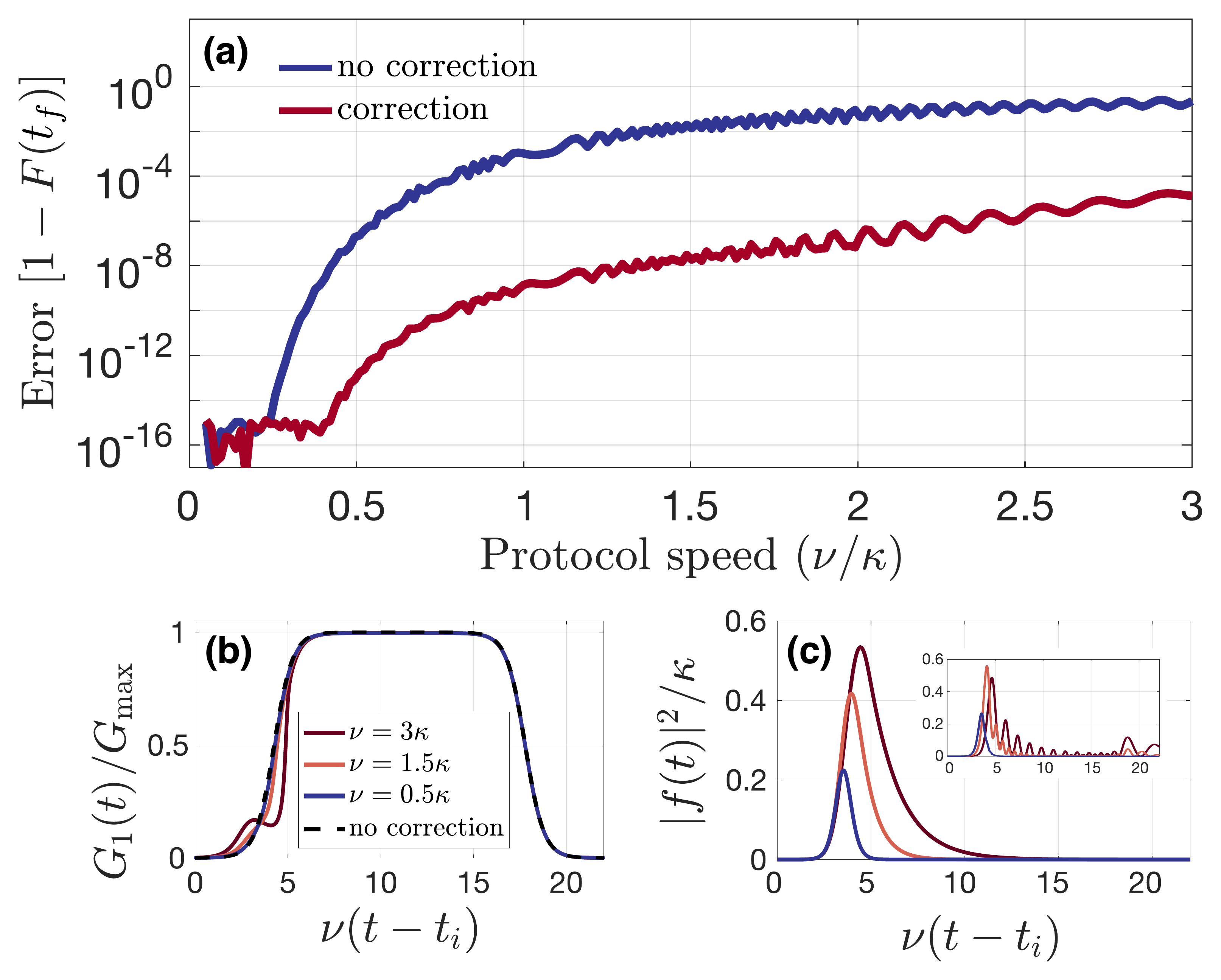}
	\end{center}
	\caption{STA correction [c.f.~Eq.~\eqref{eq:G1corrDK}] for constrained STIRAP-style state transfer, where $G_2(t)=g$ at all times.
	Corrections are based on the adiabatic control pulse given in Eq.~(\ref{eq:coupling_Duan_Kimble_1});
	we set $g=6\kappa$, and $G_{\rm max}=30 \kappa$.
	(a)  Fidelity error at the end of the protocol as a function of protocol speed $\nu$ for the uncorrected and corrected
	protocols.  The correction yields $\sim 6$ orders of magnitude improvement for a wide range of protocol speeds.    
	The fidelity error of the corrected protocol is only limited by our truncation of the control pulses  
	(initial and final time have been chosen such that $G_1(t_i)=G_1(t_f)=10^{-3}g$), and the finite amplitude $G_{\rm max}$ of the pulse at intermediate time.
	(b) Evolution of the control field $G_1(t)$ during the protocol, without (dashed line) and with (solid lines) correction;
	legend indicates value of protocol speed $\nu$.  
	(c) Shape of the emitted temporal mode in the waveguide when using the corrected pulse (the same quantity has been represented in the inset for the non-corrected pulse); the curves are for different values of $\nu$ [same
	labelling as in (b)].}
	\label{fig:Duan_Kimble_results}
\end{figure}

When $G_2=g$ is constant, the uncorrected adiabatic transfer protocol here involves slowly 
ramping $G_1(t)$ up from zero until it is $\gg g$ at a time $t \sim t_{\rm mid}$, so that the adiabatic dark state is just $\ket*{C}$.  One then waits
for a time $\sim t_0 > \kappa$ for the state to decay to the waveguide, and then ramps $G_1(t)$ back down to zero \cite{Fleischhauer2000,Duan2003}.      
A simple pulse shape that accomplishes this is (c.f.~Fig.~\ref{fig:Duan_Kimble_results}b):
\begin{align}
	G_1(t)=\frac{G_{\rm max}}{2}
		\left(	\tanh[\nu t]-\tanh[\nu (t-t_0)]	\right).
\label{eq:coupling_Duan_Kimble_1}
\end{align}
The rate $\nu$ here sets both the rate of the initial ramp and the time $t_{\rm mid}$, and $t_0$ sets the delay between the turn on and turn off phases.  This pulse would give a perfect transfer in the limit $G_{\rm max} \gg \nu, \kappa, g$.  

Our goal is make the above protocol perfect even when non-adiabatic and dissipative effects are important, i.e.~when
$\nu / G_{\rm max}, \kappa / G_{\rm max}$ are finite.  
We start by insisting that our correction does not modify the amplitude $G_2(t)=g$, which implies [c.f.~Eq.~(\ref{eqs:GCorrs})] $g_x(t)\sin\theta(t)+g_z(t)\cos\theta(t)=0$.  Using this constraint in Eq.~\eqref{eq:gx},\eqref{eq:gz} results in a differential equation for the dressing amplitude $\mu(t)$,
\begin{multline}
		\dot{\mu}(t)\sin\theta(t)\sin\mu(t)  =  
		\dot{\theta}(t)\cos\theta(t)\cos\mu(t)  -g \sin\mu(t)
			 \\
		+\frac{\kappa}{2}\sin\theta(t)\cos\mu(t)\left(1-\sin^2\theta(t)\cos^2\mu(t)\right),		
				 \label{eq:DiffEqMuDK}				
\end{multline}
and also directly links the form of the corrected pulse to $\mu(t)$:
\begin{align}
	G_{1,{\rm corr}}(t)  &=  G_1(t)+\frac{\dot{\mu}(t)-\frac{\kappa}{4}\sin^2[\theta(t)]\sin[2\mu(t)]}{\cos[\theta(t)]}.
		\label{eq:G1corrDK}
\end{align}

Finding a pulse that corrects for non-adiabatic and dissipative errors thus requires solving
Eq.~\eqref{eq:DiffEqMuDK} with the boundary condition $\mu(t_i) = 0$.  
This however is not enough: we also require that the dressing strength $\mu(t)$ vanish in the middle of the protocol (i.e.~$t \sim t_{\rm mid}$), so that the dressed dark state is just $\ket*{C}$ and can decay fully into the waveguide.  A priori, there is no guarantee that in general, the solution of Eq.~(\ref{eq:DiffEqMuDK}) [with $\mu(t_i) = 0$] fulfils this condition.  

Serendipitously, for the uncorrected pulse sequence in  Eq.~\eqref{eq:coupling_Duan_Kimble_1}, we find via explicit numerical integration of Eq.~\eqref{eq:DiffEqMuDK} that the dressing $\mu(t)$ does indeed almost completely turn off in the middle of the protocol as desired.  We use
Eqs.~\eqref{eq:DiffEqMuDK},\eqref{eq:G1corrDK} to find the corrected pulse $G_{1,{\rm corr}}(t)$ on the interval $(-\infty,t_0/2)$.  For $t > t_0/2$, the transfer is effectively complete, and it does not matter how we turn off the pulse [i.e.~there is no need to correct $G_1(t)$].  We thus have the pulse turn off exactly the same way as the uncorrected pulse, i.e. $G_{1,{\rm corr}}(t) = A
G_{1}(t)$ (where the constant $A$ is chosen to ensure continuity).

Fig.~\ref{fig:Duan_Kimble_results} shows corrected pulses and fidelity improvements resulting from this approach.  We use finite initial and final times, chosen so that that $G_1(t_i) = G_1(t_f) =10^{-3}g$, and also pick the delay time $t_0=-2t_i+5/\nu$ to scale with $1/\nu$; the result is that the total pulse duration scales inversely with the speed parameter $\nu$.  Fig.~\ref{fig:Duan_Kimble_results}(a) demonstrates an impressive six orders of magnitude suppression of the fidelity error in regimes where both adiabatic and dissipative errors contribute equally.  Note that for extremely fast pulses $\nu \gg \kappa$, both corrected and uncorrected protocols are limited by there not being enough time for the state to decay to the waveguide.  Fig.~\ref{fig:Duan_Kimble_results}(b) demonstrates that the correction to the pulses are extremely simple, corresponding to a simple ``wiggle" being added during the turn-on phase.  

Finally, our correction also has the benefit of resulting in extremely simple and smooth temporal mode shapes.  Fig.~\ref{fig:Duan_Kimble_results}(c) shows the temporal mode shapes resulting from the corrected protocol, while the inset shows the mode shapes obtained in the original, uncorrected protocol.  The fast oscillations here (which are absent when one uses the correction) would make subsequent ``catch" operations extremely difficult.  

{\it Conclusions-- } We have presented a general strategy for using STA techniques to accelerate adiabatic processes for systems which include an infinite-dimensional continuum.  Focusing on the problem of adiabatic state transfer between a discrete system and a waveguide, our technique allows one to both accelerate standard STIRAP-style adiabatic approaches {\it and} completely counteract dissipative errors generated by the coupling to the continuum. The application of this method on two experimentally relevant situations shows an improvement of the fidelity by several orders of magnitude, even when the intermediate level is subject to damping. In the future, this technique could be generalized to describe more general many-body systems, where part of the system could be modelled as an effective continuum.  \\

We acknowledge the support of the AFOSR MURI program on quantum state transfer.


\clearpage

\global\long\def\theequation{S\arabic{equation}}

\global\long\def\thefigure{S\arabic{figure}}

\setcounter{equation}{0}

\setcounter{figure}{0}

\setcounter{secnumdepth}{2}

\thispagestyle{empty}
\onecolumngrid

\begin{center}
{\fontsize{12}{12}\selectfont
\textbf{Supplemental \hspace{-0.2cm} Material for
	``Speeding \hspace{-0.2cm} up adiabatic quantum state transfer by using dressed states''\\ [5mm]}}
{\normalsize Alexandre Baksic, Ron Belyansky, Hugo Ribeiro, and Aashish A. Clerk\\[1mm]}
{\fontsize{9}{9}\selectfont
\textit{Department of Physics, McGill University, Montréal, Quebec, Canada H3A 2T8 }}
\end{center}   
\normalsize

\section{Projection of the Hamiltonian describing the systems in Fig.~1 in the one excitation subspace.}
The system in Fig.~\ref{fig:sketchSys}(b) of the main text represents three-bosonic modes linearly coupled to one another, with one of them being in addition coupled to a waveguide. This type of Hamiltonian ca be realized e.g. in quantum optomechanical systems \cite{KippenbergRMP2014Supp} by coupling two optical cavities to the same mechanical mode and coupling one of them to a waveguide. Then, by driving each of those cavities independently on their red sideband, it is possible to independently control the couplings of each cavity mode to the mechanical one via the intensity of the applied lasers. The Hamiltonian describing this system, in a frame rotating at the frequency of the two applied lasers and under the rotating wave approximation, is given by ($\hbar=1$) 
\begin{align}
\Ham_{\rm OM}(t)=&\left[G_1(t)\ha^{\phantom{\dagger}}+G_2(t)\haa\right] \hbd+\left[G_1(t)\had+G_2(t)\haad\right]\hb+\int_{-\omega_{\rm max}/2}^{\omega_{\rm max}/2}d\omega \left[\omega\, \hcd(\omega)\hc(\omega)+\sqrt{\frac{\kappa}{2\pi}}\left(\hcd(\omega)\haa+\hc(\omega)\haad\right)\right]\label{eq:Ham_OM_supp}
\end{align}
where $\hc(\omega)$ is the photon annihilation operator of a mode at frequency $\omega$ in the waveguide, obeying the commutation relation $[\hc(\omega),\hcd(\omega')]=\delta(\omega-\omega')$. Then, by defining the one excitation states
\begin{align}
\ket*{A}=\had\vac\,\,,\,\,\ket*{B}=\hbd\vac\,\,,\,\,\ket*{C}=\haad\vac\,\,,\,\,\ket*{D_\omega}=\hcd(\omega)\vac,
\end{align}
where $\vac$ represents the vacuum of the whole system (i.e. no excitation in any of the modes), the Hamiltonian Eq.~\eqref{eq:Hamiltonian_levels_full} of the main text is easily found by projecting the Hamiltonian Eq.~\eqref{eq:Ham_OM_supp} in this subspace. Moreover, since this Hamiltonian is linear, by starting in a one excitation state we are insured that the time evolution will stay in this subspace.

The system in Fig.~\ref{fig:sketchSys}(c) of the main text  (and studied in \cite{Fleischhauer2000Supp,Duan2003Supp}) represents a three-level lambda system placed inside a cavity, with one of its transition driven by a laser field, and the other transition coupled to a cavity mode $\hat{a}$ which in its turn is coupled to a waveguide. The Hamiltonian of that system can be written as
\begin{align}
\Ham_{\rm \Lambda-cav}(t)=G_1(t)\Big{(}\dyad*{e}{g}+\dyad*{g}{e}\Big{)}+g\hat{a}\Big{(}\dyad*{g}{s}+\dyad*{s}{g}\Big{)}+\int_{-\omega_{\rm max}/2}^{\omega_{\rm max}/2}d\omega \left[\omega\, \hcd(\omega)\hc(\omega)+\sqrt{\frac{\kappa}{2\pi}}\left(\hcd(\omega)\haa+\hc(\omega)\haad\right)\right].\label{eq:Ham_cav_lambda_supp}
\end{align}
Then, by defining the one excitation states
\begin{align}
\ket*{A}=\ket*{e}\otimes\vac\,\,,\,\,\ket*{B}=\ket*{g}\otimes\vac\,\,,\,\,\ket*{C}=\ket*{s}\otimes\hat{a}^\dagger\vac\,\,,\,\,\ket*{D_\omega}=\ket*{s}\otimes\hcd(\omega)\vac,
\end{align}
where $\vac$ now represents the vacuum of all the bosonic modes, the Hamiltonian Eq.~\eqref{eq:Hamiltonian_levels_full} of the main text is again easily found by projecting the Hamiltonian Eq.~\eqref{eq:Ham_cav_lambda_supp} in this subspace.

\section{Derivation of the non-Hermitian Hamiltonian $H_1(t)$}
The Schr\"odinger equation for the amplitudes $\uA(t)$, $\uB(t)$, $\uC(t)$ and $\uWG(t)$ [Eq.~\eqref{eq:WF_1exc_WG} of the main text] is given by a set of coupled differential equations
\begin{align}
&\frac{d}{dt}\uA(t)=-iG_1(t)\uB(t)\\
&\frac{d}{dt}\uB(t)=-i\left[G_1(t)\uA(t)+G_2(t)\uC(t)\right]\\
&\frac{d}{dt}\uC(t)=-i\left[G_2(t)\uB(t)+\sqrt{\frac{\kappa}{2\pi}}\int_{-\omega_{\rm max}/2}^{\omega_{\rm max}/2}d\omega\,\uWG(\omega,t)\right]\\
&\frac{d}{dt}\uWG(t)=-i\left[\omega\uWG(\omega,t)+\sqrt{\frac{\kappa}{2\pi}}\uC(t)\right]
\end{align}
We can first formally solve the differential equation for the waveguide amplitude $\uWG(t)$ as
\begin{align}
\uWG(\omega,t)=-i\sqrt{\frac{\kappa}{2\pi}}\int_{t_i}^{t}d\tau e^{-i\omega(t-\tau)}\uC(\tau),
\end{align}
where we assumed the initial condition $\uWG(\omega,t_i)=0\,\forall\,\omega$. Then, by reintroducing this solution into the other differential equations and taking the Markovian limit $\omega_{\rm max}\rightarrow\infty$, we obtain
\begin{align}
&\duA(t)=-iG_1(t) \uB(t)\nonumber\\
&\duB(t)=-iG_1(t) \uA(t)-iG_2(t) \uC(t)\nonumber\\
&\duC(t)=-iG_2(t) \uB(t)-\frac{\kappa}{2} \uC(t).
\end{align}
This set of differential equations corresponds to a Schr\"odinger equation for the effective non-Hermitian Hamiltonian
\begin{align}
\hat{H}_1(t)=\hat{H}_0(t)-i\frac{\kappa}{2}\dyad*{C}{C}\label{eq:AppdxH1},
\end{align}
and we thus see that the Markovian continuum (waveguide) effectively acts on the state $\ket*{C}$ as a damping term.

\section{Equations of motions for bosonic modes}
In this section we want to emphasize that, thanks to the linearity of the systems considered in this paper, the Schr\"odinger equation in the one excitation subspace [as described by the non-Hermitian Hamiltonian Eq.~\eqref{eq:AppdxH1}] is equivalent to the Heisenberg equations of motion for the bosonic modes $\ha$, $\hb$, and $\hc$ that appear in Hamiltonian Eq.~\eqref{eq:Ham_OM_supp}. This means that the protocols described in the main text can implement a fast and efficient conversion of an arbitrary states in mode $\ha$ to temporal modes in the waveguide (not just an single excitation Fock state). The equations of motion for the bosonic modes are found by computing $\dot{\hat{f}}=-i[\hat{f},\Ham_{\rm OM}]$ (where $\hat{f}=\ha,\hb,\haa,\hc(\omega)$) i.e.
\begin{align}
&\frac{d}{dt}\ha=-iG_1(t)\hb\\
&\frac{d}{dt}\hb=-i\left(G_1(t)\ha+G_2(t)\haa\right)\\
&\frac{d}{dt}\haa=-i\left(G_2(t)\hb+\sqrt{\frac{\kappa}{2\pi}}\int_{-\omega_{\rm max}/2}^{\omega_{\rm max}/2}d\omega\hc(\omega)\right)\\
&\frac{d}{dt}\hc(\omega)=-i\left(\omega\hc(\omega)+\sqrt{\frac{\kappa}{2\pi}}\haa\right).
\end{align}
By formally solving the equation for the waveguide modes $\hc(\omega)$as,
\begin{align}
\hc(\omega,t)=-i\sqrt{\frac{\kappa}{2\pi}}\int_{t_i}^tdt'\exp[-i\omega(t-t')]\haa(t')+\exp[-i\omega(t-t_i)]\hc(\omega,t_i),
\end{align}
introducing the input mode
\begin{align}
\hc_{\rm in}(t)=\frac{-i}{\sqrt{2\pi}}\int_{-\omega_{\rm max}/2}^{\omega_{\rm max}/2}\,d\omega\,\exp[-i\omega(t-t_i)]\hc(\omega,t_i)
\end{align}
and taking the Markovian limit $\omega_{\rm max}\rightarrow\infty$, those equations can be reduced to
\begin{align}
&\frac{d}{dt}\ha(t)=-iG_1(t)\hb(t)\\
&\frac{d}{dt}\hb(t)=-i\left[G_1(t)\ha(t)+G_2(t)\haa(t)\right]\\
&\frac{d}{dt}\haa(t)=-iG_2(t)\hb(t)-\frac{\kappa}{2}\haa(t)+\sqrt{\kappa}\hc_{\rm in}(t).
\end{align}
Theses can be cast in the form
\begin{align}
\frac{d}{dt}\vec{\hat{a}}(t)=-iH_1(t)\vec{\hat{a}}(t)+\vec{\hat{\zeta}}(t)
\end{align}
 where we introduced $\vec{\hat{a}}(t)=[\ha(t),\hb(t),\haa(t)]^T$, $\vec{\hat{\zeta}}(t)=[0,0,\sqrt{\kappa}\hc_{\rm in}(t)]^T$ and 
 \begin{align}
 H_1(t)=\left(\begin{array}{ccc}0&G_1(t)&0\\G_1(t)&0&G_2(t)\\0&G_2(t)&-i\kappa/2 \end{array}\right),
 \end{align}

which is the matrix representative of the Hamiltonian operator $\hat{H}_1$ Eq.~\eqref{eq:AppdxH1}. The formal solution to these equations can be written as
\begin{align}
\vec{\hat{a}}(t)=U(t,t_i)\vec{\hat{a}}(t_i)+\int_{t_i}^t\,dt'\,U(t,t')\vec{\hat{\zeta}}(t'),
\end{align}
where $U(t,t')$ is the evolution operator that obeys the Schr\"odinger equation
\begin{align}
\frac{d}{dt}U(t,t')=-iH_1(t)U(t,t')
\end{align}
with the initial condition $U(t',t')=\mathbb{1}$. Thus, if the input field $\hc_{\rm in}(t)$ is in the vacuum, it will not contribute to any normally ordered correlation function involving a component of $\vec{\hat{a}}(t)$. This shows the correspondence between the Schr\"odinger equations for the amplitudes $\uA(t)$, $\uB(t)$, $\uC(t)$ and the Heisenberg equations of motion for the bosonic operators $\ha(t)$, $\hb(t)$, $\haa(t)$, and thus that the corrected protocols obtained in the main text can also be used to implement a fast and efficient conversion of arbitrary states in mode $\ha$ to temporal modes in the waveguide.

\section{Derivation of the time dependent controls $g_x(t)$ and $g_z(t)$}
In this section we show in detail how to choose the additional controls $g_x(t)$ and $g_x(t)$ to cancel leakage out of the dressed dark states. By using the dressing and the correction defined in the main text we can obtain the following expression for the corrected Hamiltonian in the dressed state basis [Eq.~\eqref{eq:Hsad} of the main text]:
\begin{align}
\Hsad=&\Usad^{\dagger}(t)[\Ham_{{1,\rm ad}}(t)+\Uad^\dagger(t)\Ham_{{\rm cor}}(t)\Uad(t)]\Usad(t)-i\Usad^{\dagger}(t)\dUsad(t)\\
	=&[\cos\mu(G_0+g_z)+\dot{\theta}\sin\mu]\left(\dyad*{\widetilde{+}}{\widetilde{+}}-\dyad*{\widetilde{-}}{\widetilde{-}}\right)-i\frac{\kappa}{4}\left[1-\sin^2\theta\cos^2\mu\right]\left(\dyad*{\widetilde{+}}{\widetilde{+}}+\dyad*{\widetilde{-}}{\widetilde{-}}\right)\nonumber\\
	&-i\frac{\kappa}{2}\cos^2\theta\sin^2\mu\dyad*{\widetilde{\dark}}{\widetilde{\dark}}+\frac{\kappa}{2}\sin(2\theta)\sin\mu\left(\dyad*{\widetilde{-}}{\widetilde{+}}-\dyad*{\widetilde{+}}{\widetilde{-}}\right)-i\frac{\kappa}{4}\left[\cos^2\theta-\sin^2\theta\sin^2\mu\right]\left(\dyad*{\widetilde{-}}{\widetilde{+}}+\dyad*{\widetilde{+}}{\widetilde{-}}\right)\nonumber\\
	&+\left[\dot{\mu}+g_x-\frac{\kappa}{4}\sin^2\theta\sin(2\mu)\right]\left(\frac{\ket*{\widetilde{+}}-\ket*{\widetilde{-}}}{\sqrt{2}}\bra*{\widetilde{\dark}}\right)+i\left[\left\{G_0+g_z\right\}\sin\mu-\left\{\dot{\theta}+\frac{\kappa}{4}\sin(2\theta)\right\}\cos\mu\right]\left(\frac{\ket*{\widetilde{+}}+\ket*{\tilde{-}}}{\sqrt{2}}\bra*{\tilde{\dark}}\right)\nonumber\\
	&+\left[\dot{\mu}+g_x+\frac{\kappa}{4}\sin^2\theta\sin(2\mu)\right]\left(\ket*{\widetilde{\dark}}\frac{\bra*{\widetilde{+}}-\bra*{\widetilde{-}}}{\sqrt{2}}\right)+i\left[-\left\{G_0+g_z\right\}\sin\mu+\left\{\dot{\theta}-\frac{\kappa}{4}\sin(2\theta)\right\}\cos\mu\right]\left(\ket*{\widetilde{\dark}}\frac{\bra*{\widetilde{+}}+\bra*{\widetilde{-}}}{\sqrt{2}}\right)
\end{align}
and we thus see that in order to cancel the terms that are moving population from the dressed dark state to $\ket*{\tilde{\pm}}$, (i.e. all the matrix elements of the form $\dyad*{\widetilde{j}}{\widetilde{\dark}}$ with $j=+,-$), we need the two following equations to be satisfied
\begin{align}
&\dot{\mu}+g_x-\frac{\kappa}{4}\sin^2\theta\sin(2\mu)=0\\
&\left\{G_0+g_z\right\}\sin\mu+\left\{\dot{\theta}+\frac{\kappa}{4}\sin(2\theta)\right\}\cos\mu=0,
\end{align}
which is done by choosing the two controls
\begin{align}
&g_x=-\dot{\mu}+\frac{\kappa}{4}\sin^2\theta\sin(2\mu)\\
&g_z=\frac{1}{\tan\mu}\left[\dot{\theta}+\frac{\kappa}{4}\sin(2\theta)\right]-G_0.
\end{align}

\section{plots of the dressing strength $\mu(t)$ in the case where $G_2(t)=g,\,\forall\,t$ }

\begin{figure}[t]
	\begin{center}
		\includegraphics[width=0.5 \columnwidth]{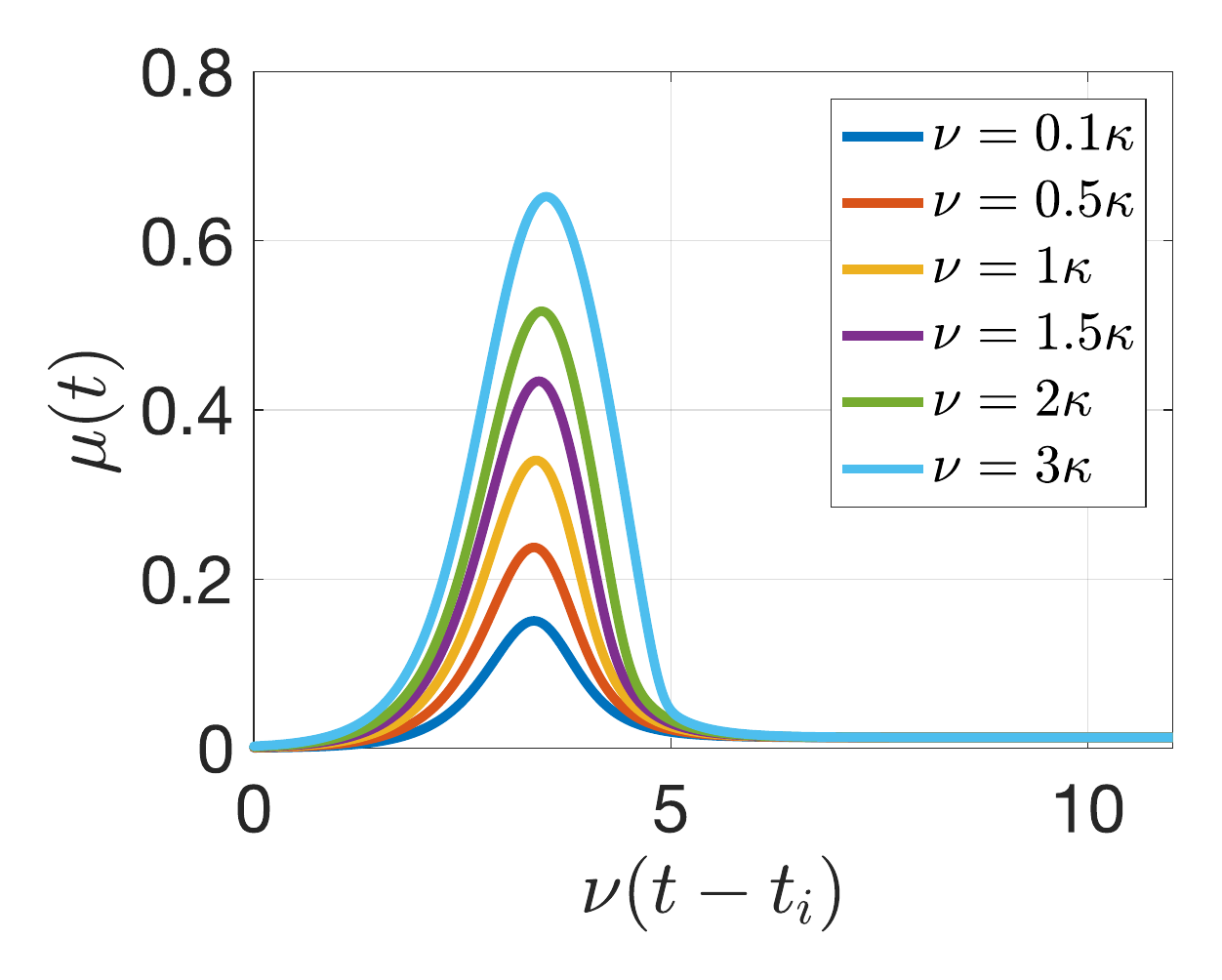}
	\end{center}
	\caption{Dressing strength $\mu(t)$ from $t_i$ to $t_0/2$, found by solving the differential equation in the main text for different values of the protocol speed $\nu$. We set $g=6\kappa$ and $G_{\rm max}=30\kappa$.
	}

	\label{fig:EPAPS_Mu}
\end{figure}
In Fig.~\ref{fig:EPAPS_Mu} we plot the dressing strength $\mu(t)$ found by solving the differential equation [Eq.~\eqref{eq:DiffEqMuDK} of the main text]  from $t_i$ [with inital value $\mu(t_i)=0$] to $t_0/2$. We see that it behaves as expected and goes from 0 at initial time $t_i$ to almost $0$ ($\approx0.013$) at intermediate time $t_0/2$. The fact that the dressing does not totally turn off at $t_0/2$ is a consequence of our choice of initial protocol [Eq.~\eqref{eq:coupling_Duan_Kimble_1} of the main text] for which the angle $\theta(t)$ is not exactly $\pi/2$ at time $t_f$ but rather $\arctan(G_{\rm max}/g)$, which is also the reason why we need to introduce the scaling factor $A$ in the main text. We would need $G_{\rm max}\rightarrow\infty$ for the dressing to totally vanish at time $t_0/2$ which would not be experimentally realistic.

\end{document}